\documentclass[%
 reprint,
 superscriptaddress,
amsmath,amssymb,
aps,
pra,
]{revtex4-2}
\usepackage{graphicx}
\usepackage{subfigure}
\usepackage{float}
\usepackage{dcolumn}
\usepackage{bm}
\usepackage{rotating}
\usepackage{color}
\usepackage{xcolor}
\usepackage{braket}
\usepackage{float}
\usepackage[colorlinks,citecolor=blue,urlcolor=blue,linkcolor=blue]{hyperref}
\begin{document}

\preprint{APS/123-QED}


\title{Dimer-driven multiple reentrant localization with composite potential}

\author{Pei-Jie Chang}
    \affiliation{State Key Laboratory of Low-Dimensional Quantum Physics and Department of Physics, Tsinghua University, Beijing 100084, China}

\author{Mu-Xi Zheng}
    \affiliation{State Key Laboratory of Low-Dimensional Quantum Physics and Department of Physics, Tsinghua University, Beijing 100084, China}

\author{Quan-Feng Lu}
    \affiliation{State Key Laboratory of Low-Dimensional Quantum Physics and Department of Physics, Tsinghua University, Beijing 100084, China}
    \affiliation{Beijing Academy of Quantum Information Sciences, Beijing 100193, China}

\author{Dong Ruan}
    \affiliation{State Key Laboratory of Low-Dimensional Quantum Physics and Department of Physics, Tsinghua University, Beijing 100084, China}
    \affiliation{Frontier Science Center for Quantum Information, Tsinghua University, Beijing 100084, China}

\author{Gui-Lu Long}
    \email{gllong@tsinghua.edu.cn}
    \affiliation{State Key Laboratory of Low-Dimensional Quantum Physics and Department of Physics, Tsinghua University, Beijing 100084, China}
    \affiliation{Frontier Science Center for Quantum Information, Tsinghua University, Beijing 100084, China}
    \affiliation{Beijing Academy of Quantum Information Sciences, Beijing 100193, China}
    \affiliation{Beijing National Research Center for Information Science and Technology, Beijing 100084, China}

\begin{abstract}


Recent studies have revealed reentrant localization transitions in quasiperiodic one-dimensional lattices, where the competition between dimerized hopping and staggered disorder plays a central role. However, it remains unclear under what conditions reentrant localization and in particular multiple reentrant localization can occur. Here we investigate localization phenomena in a one-dimensional lattice subject to a periodic potential and an additional quasiperiodic modulation. Using both eigenstate-based indicators and experimentally accessible dynamical observables, we identify robust reentrant multiple reentrant localization transitions. We show that these transitions are uniquely stabilized by the dimer structure of the unit cell, where the competition between the onsite periodic potential and the quasiperiodic modulation becomes most pronounced. By systematically varying the periodicity parameter $\alpha$ and the quasiperiodic frequency $\beta$, we find that the robust multiple reentrant localization behavior disappears for any deviation from the dimer configuration, confirming its essential role. Our results suggest that the interplay between these competing factors drives the multiple reentrant localization transitions. 

\end{abstract}

\maketitle


\section{Introduction}

Anderson localization is one of the central topics in condensed matter physics \cite{PhysRev.109.1492, RevModPhys.57.287, RevModPhys.80.1355}. It arises from the coherent interference of scattered wave functions induced by disorder, leading to the localization of electronic states in periodic lattices. This phenomenon has been experimentally observed in a wide variety of physical systems \cite{roati2008anderson, PhysRevLett.103.013901, PhysRevLett.122.100403, maynard2001acoustical, hu2008localization, yamilov2023anderson, segev2013anderson, mookherjea2014electronic, vatnik2017anderson, PhysRevLett.100.013906, ghasempour2023anderson, PhysRevA.89.022309}. According to scaling theory \cite{PhysRevLett.42.673, PhysRevLett.47.882}, in one- and two-dimensional (1D and 2D) systems with random disorder, any finite disorder strength localizes all eigenstates, so that neither a mobility edge nor an extended-to-localized transition occurs. In contrast, three-dimensional (3D) systems can undergo a disorder-driven transition from extended to localized states, characterized by a well-defined mobility edge that separates the two regimes \cite{PhysRevLett.134.053601, N_Mott_1987, PhysRevLett.61.2144}. The Aubry–André–Harper (AAH) model provides a paradigmatic exception: although it is a one-dimensional system, the presence of a quasiperiodic potential with an irrational modulation wave number enables an extended-to-localized transition at finite disorder strength \cite{AAH, PhysRevLett.65.88}. However, due to its intrinsic self-duality, this transition occurs without the emergence of a mobility edge \cite{AAH, PhysRevB.96.054202}. Various extensions of the AAH model, such as incorporating long-range hopping with power law decay \cite{kim2010quantum, richerme2014non, britton2012engineered, islam2013emergence, schauss2015crystallization, PhysRevLett.121.093602, PhysRevB.103.075124, roy2021entanglement, PhysRevB.110.094203, PhysRevResearch.3.013148}, mosaic-type disorder \cite{PhysRevLett.125.196604, PhysRevB.108.054204, PhysRevB.104.064203}, further enrich the physics by producing mobility edges \cite{PhysRevLett.114.146601, PhysRevB.96.085119}, critical states \cite{PhysRevLett.125.073204, PhysRevB.93.104504}, and non-trivial topological phase transitions \cite{PhysRevLett.110.176403, PhysRevB.106.224505} in low-dimensional systems. 

Typically, either in systems with random disorder or in those with quasiperiodic disorder, increasing the strength of the disorder monotonically drives at most one single localization transition \cite{AAH, PhysRevLett.114.146601, PhysRevB.96.085119}. However, recent studies on one-dimensional quasiperiodic Su–Schrieffer–Heeger (SSH) systems have revealed an unexpected reentrant localization behavior: as the strength of staggered quasiperiodic disorder increases monotonically, some eigenstates evolve from extended to localized, then delocalized again, and finally become localized \cite{PhysRevLett.126.106803}. The phenomenon of reentrant localization has been extensively validated in subsequent studies \cite{PhysRevB.106.054307, PhysRevB.105.054204, PhysRevB.108.L100201, PhysRevB.105.L220201, wu2021non, PhysRevB.110.134203, PhysRevB.107.224201, PhysRevB.107.035402, ganguly2024phenomenon, li2023multiple, PhysRevA.106.053312, PhysRevA.109.043308, lu2023robust, PhysRevB.109.195427, PhysRevB.110.184208, sarkar2024signature, gong2021comment}. And this reentrant localization has been shown to exhibit universal critical exponents at the different transition points \cite{PhysRevB.105.214203}. Further investigations have demonstrated that reentrant localization can also arise in diverse contexts, such as composite-potential systems \cite{PhysRevB.105.L220201, PhysRevB.107.224201, PhysRevB.111.134204}, models with long-range hopping \cite{PhysRevB.107.075128, Chang_2025}, random-disorder systems \cite{PhysRevA.106.013305, xu2024observation}, higher-dimensional lattices \cite{10.21468/SciPostPhys.13.5.116, PhysRevB.63.214202}, spinful models \cite{PhysRevA.108.033305, Sarkar_2025}, and even non-Hermitian systems \cite{PhysRevB.105.054204, PhysRevB.109.L020203, PhysRevA.108.033305, jiang2021mobility, bd1n-dclq}. In parallel with these theoretical advances, the experimental observation of reentrant localization has progressed rapidly, with signatures reported in SSH optical lattices \cite{xu2024observation}, photonic crystals \cite{PhysRevResearch.5.033170}, and polaritonic systems \cite{goblot2020emergence}. Moreover, in systems with combined periodic and quasiperiodic potentials, multiple reentrant localization transitions have been observed as the potential or disorder strength is varied \cite{PhysRevB.105.L220201, 10.21468/SciPostPhysCore.8.1.012}. Although these findings significantly deepen our understanding of localization phenomena, the microscopic origins of localization and, in particular, the conditions necessary for multiple reentrant localization transitions with changing strength of quasiperiodic disorder are still not well understood and have not yet been systematically explored.

In this work, we address these open questions by systematically exploring a one-dimensional model that combines a periodic potential with a quasiperiodic modulation \cite{10.21468/SciPostPhysCore.8.1.012}. Using eigenstate-based measures, the averaged scaling parameter and dynamical observables, we construct a comprehensive picture of the localization behavior in this system. We show that multiple reentrant localization transitions emerge only under the dimer condition with $\alpha = 1/2$, where $\alpha$ denotes the frequency of the periodic potential, and we elucidate how perturbations that preserve or break the dimer structure, respectively, stabilize or destroy this phenomenon. Furthermore, by varying $\alpha$ and the quasiperiodic frequency $\beta$, we map the parameter regimes where multiple reentrant localization persists and clarify the crucial role of the unit-cell symmetry. Taken together, our study not only identifies the microscopic origin of multiple reentrant localization transitions in quasiperiodic lattices but also offers guidance for engineering and controlling multiple reentrant phenomena. 

The remainder of this paper is organized as follows. In Sec.~\ref{se2}, we introduce the Hamiltonian model and describe the computational methods employed in our analysis. In Sec.~\ref{se3}, we present our results on multiple reentrant localization transitions, as characterized by both eigenstate-based measures and dynamical observables. We further investigate the effects of varying the initial phases and applying different types of perturbations, and we highlight the essential role of the dimer unit cell in sustaining this phenomenon. Finally, Sec.~\ref{se4} summarizes our conclusions. 

\section{MODEL AND APPROACH}\label{se2}

We consider a one-dimensional lattice model that includes a periodic potential and a quasiperiodic disorder term. The Hamiltonian of this model is expressed as
\begin{equation}
	\begin{aligned}
		\hat{\mathcal{H}}= & -J \sum_{i=1}^L\left(\hat{c}_i^{\dagger} \hat{c}_{i+1}+\text { h.c. }\right) \\
			& +V \sum_{i=1}^L \cos \left(2 \pi \alpha i+\phi_v\right) \hat{n}_i+\lambda \sum_{i=1}^L \cos \left(2 \pi \beta i+\phi_\lambda\right) \hat{n}_i,
	\end{aligned}
	\label{model}
\end{equation}
where $L$ represents the total number of lattice sites, $\hat{c}_i$ represents the fermionic annihilation operator of the $i$ site and operator $\hat{n}_{i}=\hat{c}_i^{\dagger}\hat{c}_i$ represents the particle number operator for the corresponding lattice site $i$. $J$ denotes the hopping strength between the nearest neighbors. For convenience, we set $J=1$ as the unit of the energy scale throughout our study. $V$ denotes the strength of the periodic potential, and $\alpha$ represents the corresponding frequency of the periodic potential. To ensure that the characteristics of the periodic potential are properly reflected in the calculations, $\alpha$ is chosen as simple integer fractions, such as $1/2$, $1/3$, $1/4$, etc., and $\phi_v$ is the initial phase of the periodic potential. $\lambda$ denotes the strength of the quasiperiodic disorder term, and $\beta$ corresponds to its associated frequency. The quasiperiodic disorder potential is implemented by selecting $\beta$ as an irrational number to ensure incommensurability. To minimize finite-size effects, we perform simulations on systems with sizes up to $28657$ sites. 

When $\lambda = 0$, the model reduces to the conventional periodic lattice model, whose eigenstates satisfy Bloch’s theorem and take the form of periodic modulated plane waves. When $V = 0$, the model corresponds to the classical AAH model. Due to its self-duality, the AAH model exhibits a transition from completely extended states to completely localized states at $\lambda_{\mathrm{AA}} = 2$ \cite{AAH}. When both $V$ and $\lambda$ are non-zero, the model reveals richer behaviors: a reentrant localization phenomenon emerges as the strength of the quasiperiodic disorder $\lambda$ increases, while multiple reentrant localization transitions occur as the strength of the periodic potential $V$ increases \cite{PhysRevB.105.L220201, 10.21468/SciPostPhysCore.8.1.012, PhysRevB.111.134204}. 

Within the single-particle lattice representation, we diagonalize the Hamiltonian in Eq.~\ref{model} to obtain the corresponding eigenenergies and eigenstates under open boundary conditions (OBCs). The localized nature of the system is entirely determined by these eigenstates. 
\begin{equation}
	\ket{\psi_m} = \sum_{i=1}^{L} \phi_i^{(m)}\ket{i}.
\end{equation}
Here, $\phi_i^{(m)}$ denotes the probability amplitude at site $i$ in the $m_{th}$ eigenstate. We characterize the localization properties of the eigenstates by calculating the inverse participation ratio ($\mathcal{I}$) and the normalized participation ratio ($\mathcal{N}$). For the $m_{th}$ eigenstate, the $\mathcal{I}$ and $\mathcal{N}$ are defined as follows:
\begin{equation}
	\begin{aligned}
		& \mathcal{I}_{(m)} = \sum_{i=1}^{L}|\phi_i^{(m)}|^4, \\
		& \mathcal{N}_{(m)} = \left(L\sum_{i=1}^{L}|\phi_i^{(m)}|^4\right)^{-1}.
	\end{aligned}
\end{equation}

For extended states, the value of $\mathcal{I}_{(m)}$ scales as $1/L$. Since $L$ is typically large in our calculations, $\mathcal{I}_{(m)}$ of extended states approaches zero. In contrast, for localized states, $\mathcal{I}_{(m)}$ remains on the order of unity. The $\mathcal{N}_{(m)}$ behaves oppositely: it approaches unity for extended states, while for localized states it scales as $1/L$ and thus vanishes as $L$ increases. In a critical regime, the system exhibits a coexistence of eigenstates whose $\mathcal{I}_{(m)}$ and $\mathcal{N}_{(m)}$ remain finite and do not vanish with increasing $L$. To facilitate a more convenient characterization of the general localization properties, we introduce the mean inverse participation ratio ($\mathcal{I}$) and the mean normalized participation ratio ($\mathcal{N}$), defined as
\begin{equation}
	\begin{aligned}
		& \mathcal{I} = \frac{1}{m_2-m_1}\sum_{m=m_1}^{m_2}\mathcal{I}_{(m)}, \\
		& \mathcal{N} = \frac{1}{m_2-m_1}\sum_{m=m_1}^{m_2}\mathcal{N}_{(m)}.
	\end{aligned}
	\label{IPRNPR}
\end{equation}

To more clearly determine whether the system is in the critical regime, we compute the parameter $\eta$ consistent with \cite{PhysRevLett.126.106803, PhysRevB.101.064203}, which is defined as
\begin{equation}
	\eta=log_{10}(\mathcal{I} \times \mathcal{N}), 
\end{equation}
where $\mathcal{I}$ and $\mathcal{N}$ are averaged over all eigenstates as shown in Eq.~\ref{IPRNPR}. If the system is entirely in the extended (localized) phase, $\mathcal{I}\ (\mathcal{N})$ scales as $1/L$, while $\mathcal{N}\ (\mathcal{I})$ remains finite, leading to $\eta < -3$ in our calculations. If some eigenstates are extended while others are localized, the computed value of $\eta$ is approximately $-2.5 \sim- 1$. 

\section{RESULTS}\label{se3}

\begin{figure}
     \raggedright
     \includegraphics[width=0.5\textwidth]{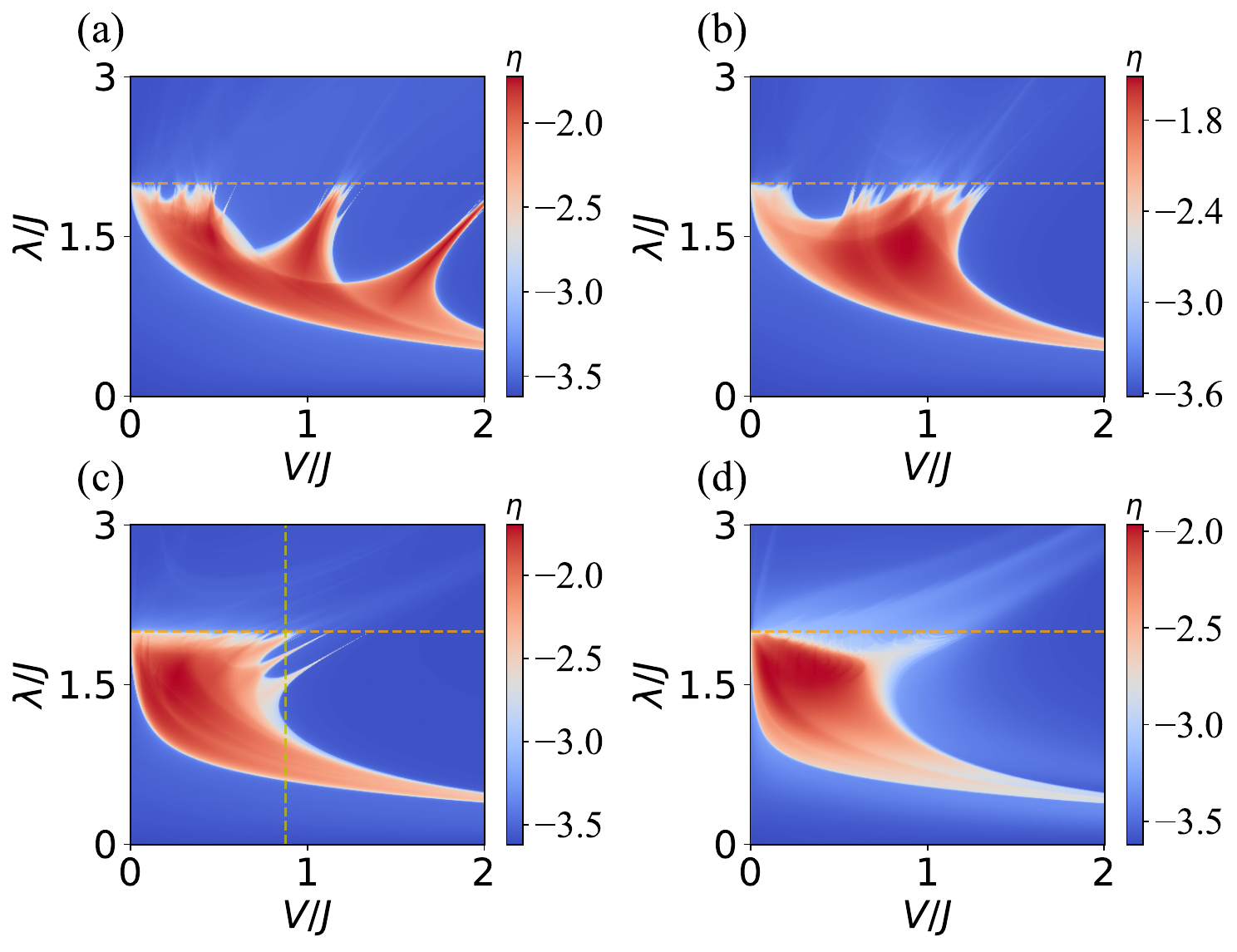}
     \caption{$\eta$ phase diagrams of the system in periodic potential strength $V/J$ and quasiperiodic disorder strength $\lambda/J$ plane. (a) $\beta=\frac{\sqrt{5}-1}{2}$, (b) $\beta=\frac{\sqrt{5}-1}{4}$, (c) $\beta=\frac{\sqrt{5}-1}{8}$, (d) $\beta=\frac{\sqrt{5}-1}{10}$, where the system size $L=4181$, $\phi_v=\phi_{\lambda}=0$ and $\alpha=1/2$. The red region corresponds to the intermediate regime in which single-particle mobility edges emerge, while the blue region denotes the parameter space where the system is either fully localized or fully extended. The orange dashed line represents the transition point of the standard AAH model, $\lambda_{\mathrm{AA}}/J = 2$. In panel (c), the yellow dashed line corresponds to periodic potential strength $V/J = 0.875$, where pronounced signatures of multiple reentrant localization can be clearly observed.}
     \label{fig1}
\end{figure}

\begin{figure}
     \centering
     \includegraphics[width=0.5\textwidth]{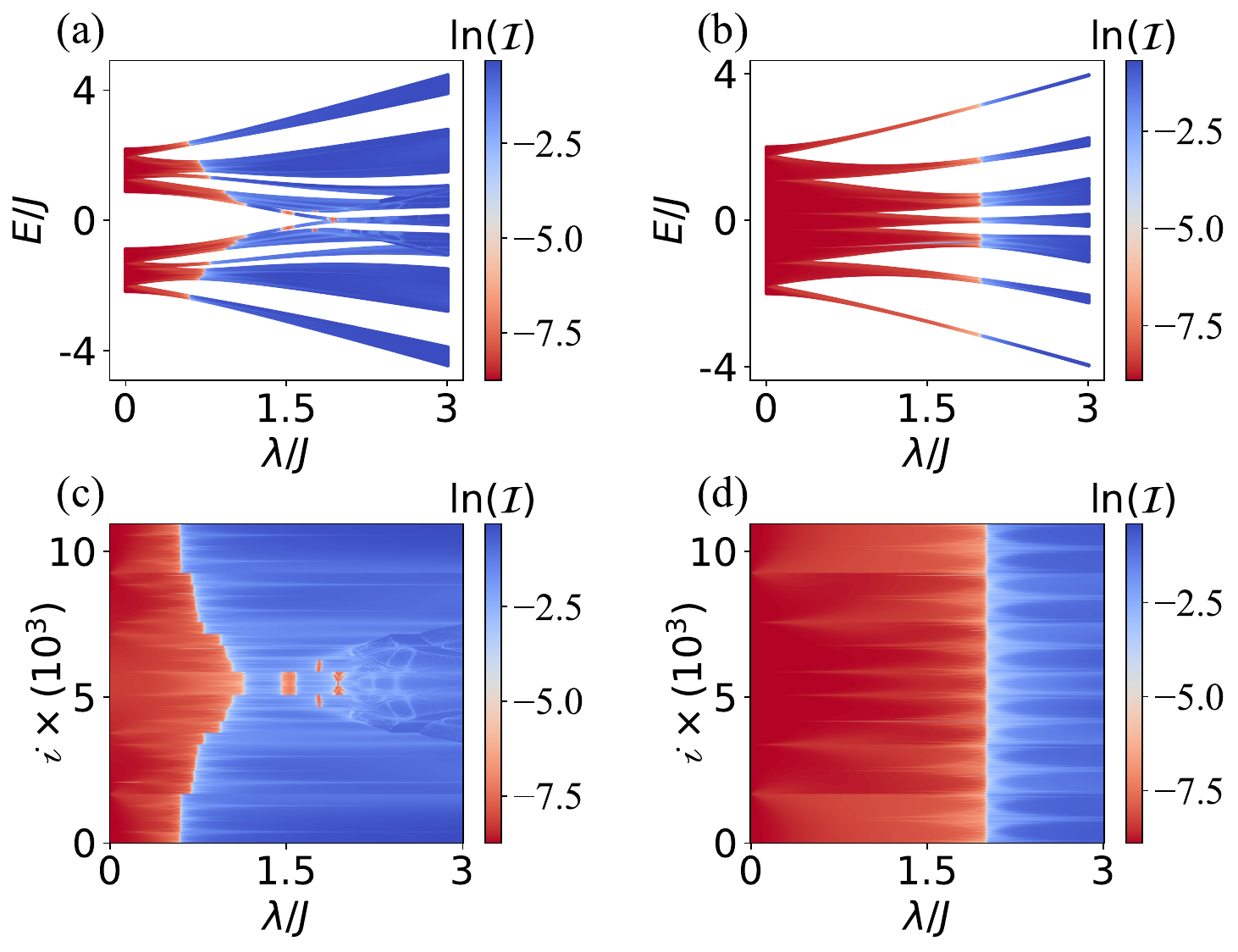}
     \caption{The $\ln(\mathcal{I})$ spectra as a function of quasiperiodic disorder strength $\lambda/J$ for (a) and (c) $\phi_v=0$, (b) and (d) $\phi_v=\pi/2$. Panels (a) and (b) display the eigenenergies on the vertical axis, while panels (c) and (d) show the indices of the corresponding eigenstates. System size $L=10946$, periodic potential strength $V/J=0.875$, $\alpha=1/2$, $\beta=\frac{\sqrt{5}-1}{8}$ and $\phi_{\lambda}=0$. The color represents the values of $\ln(\mathcal{I})$: warmer hues (towards red) indicate more extended eigenstates, whereas cooler hues (towards blue) correspond to more localized eigenstates. }
     \label{fig2}
\end{figure}

Previous literature has shown that in the SSH model, introducing a sublattice-staggered disorder with a spatial frequency $\beta$ chosen as the inverse of the golden ratio can lead to a reentrant localization phenomenon under appropriate disorder strength and suitable intra-cell and inter-cell hopping ratios \cite{PhysRevLett.126.106803}. Subsequently, even without introducing two different hoppings, simply by adding a composite potential with the frequency of the periodic potential $\alpha = 1/2$ \cite{PhysRevB.105.L220201, PhysRevB.111.134204, PhysRevB.109.L020203, PhysRevB.105.054204}, the phenomenon of reentrant localization can still be observed under appropriate parameters, as illustrated in Fig.~\ref{fig1}(a), where increasing the strength of the disorder $\lambda$ induces the reentrant localization. More interestingly, by varying the strength of the periodic potential $V$, multiple reentrant localization transitions can be observed at suitable disorder strengths $\lambda$. 

If the period of the quasiperiodic disorder gradually reduces, causing $\beta$ to deviate from $\beta = (\sqrt{5}-1)/2$, the reentrant localization phenomenon observed as a function of the disorder strength gradually weakens, as shown in Fig.~\ref{fig1}(b) for $\beta = (\sqrt{5}-1)/4$. However, when $\beta = (\sqrt{5}-1)/8$ as shown in Fig.~\ref{fig1}(c), consistent with Ref. \cite{10.21468/SciPostPhysCore.8.1.012} but without considering the power-law decaying hopping, the system exhibits multiple reentrant localization transitions as $\lambda$ increases; that is, by monotonically increasing the strength of quasiperiodic disorder — rather than the strength of the periodic potential as in previous studies \cite{PhysRevB.105.L220201, PhysRevB.105.054204}—parts of the eigenstates switch back and forth between localized and delocalized states. Upon further reduction of $\beta$ to $(\sqrt{5}-1)/10$, multiple reentrant localization transitions are no longer observed, as shown in Fig.~\ref{fig1}(d). Interestingly, regardless of the frequency choice $\beta$, once the disorder strength exceeds the transition point of the standard AAH model, $\lambda_{\mathrm{AA}} = 2$, all eigenstates become localized. No further localization transitions occur with variations in $V$ or $\lambda$, as indicated by the horizontal orange dashed line in Fig.~\ref{fig1}.

To more clearly observe the multiple reentrant localization phenomena, we fix $V = 0.875$ for direct calculations, as indicated by the yellow dashed line in Fig.~\ref{fig1}(c). The results of $\ln(\mathcal{I})$ are shown in Fig.~\ref{fig2}. When the initial phase of the periodic potential is chosen as $\phi_v = 0$, multiple reentrant localization transitions can be clearly identified in Fig.~\ref{fig2}(a) and Fig.~\ref{fig2}(c). In contrast, when the initial phase of the periodic potential is set to $\pi/2$, as shown in Fig.~\ref{fig2}(b) and Fig.~\ref{fig2}(d), no reentrant localization behavior is observed in the results of $\ln(\mathcal{I})$. In this case, the model reduces to the standard AAH model, where only a single transition point at $\lambda = 2$ exists, dictated by the self-duality condition. These results demonstrate that the initial phase of the periodic potential plays a significant role in the localization properties of the system.

\begin{figure}
     \raggedright
     \includegraphics[width=0.48\textwidth]{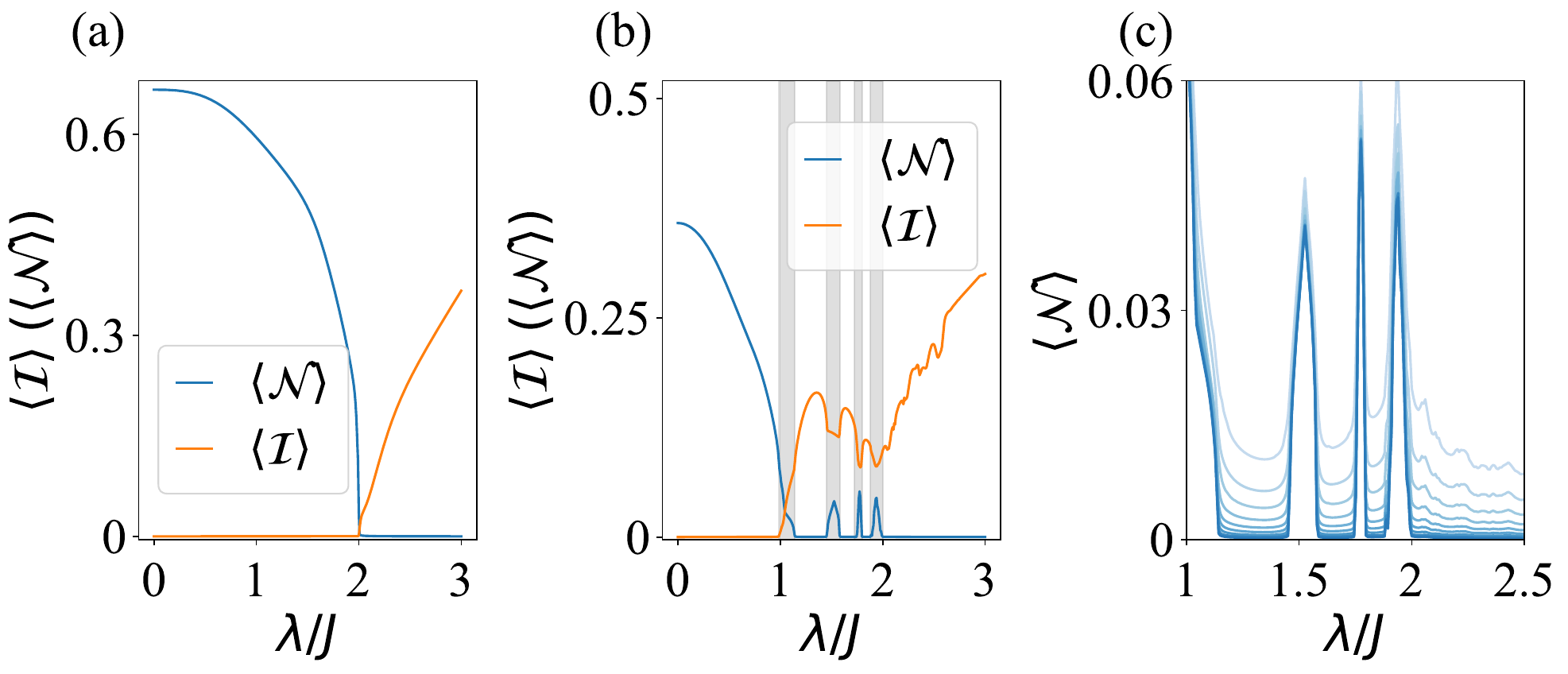}
     \caption{(a) and (b) the averaged IPR ($\mathcal{I}$) and NPR ($\mathcal{N}$), calculated for eigenstates with fractional indices $i/L \in [0.4, 0.6]$ for $\phi_{v}=\pi/2$ and $\phi_{v}=0$, respectively, for the case of periodic potential strength $V/J=0.875$ and system size $L=28657$. The shaded regions represent the critical area in panel (b), where both localized and extended states coexist. (c) The average NPR for eigenstates within fractional indices $i/L \in [0.4, 0.6]$ for $L=610, 987, 1597, 2584, 4181, 6765, 10946, 17711$ and $28657$, with the color intensity ranging from light to dark. }
     \label{fig3}
\end{figure}

In order to better characterize the range of disorder strengths over which the localization transition occurs, in Fig.~\ref{fig3}(a) and Fig.~\ref{fig3}(b), we compute the mean IPR $\mathcal{I}$ and NPR $\mathcal{N}$ using eigenstates within the interval $i/L$ $\in$ $[0.4,0.6]$. In Fig.~\ref{fig3}(a), where the initial phase of the periodic potential is set to $\pi/2$, the model reduces to the standard AAH model and thus no intermediate critical states appear. In Fig.~\ref{fig3}(b), with the initial phase set to $0$, four intervals of the quasiperiodic disorder strength can be observed that host critical states. Among them, $1.45 \sim 1.58$, $1.74 \sim 1.80$, and $1.88 \sim 2.00$ correspond to the reentrant localization phenomena induced by the increase in the strength of disorder, which are the focus of our study. To eliminate possible finite-size effects, we further calculate the mean NPR $\mathcal{N}$ for the system sizes $L = 610, 987, 1597, 2584, 4181, 6765, 10946, 17711$ and $28657$, as shown in Fig.~\ref{fig3}(c). The color gradient from light to dark represents an increase in system size. It is evident that the multiple reentrant localization phenomena persist with increasing $L$, indicating that the effect is robust and does not vanish in the thermodynamic limit. 

\begin{figure}
     \centering
     \includegraphics[width=0.48\textwidth]{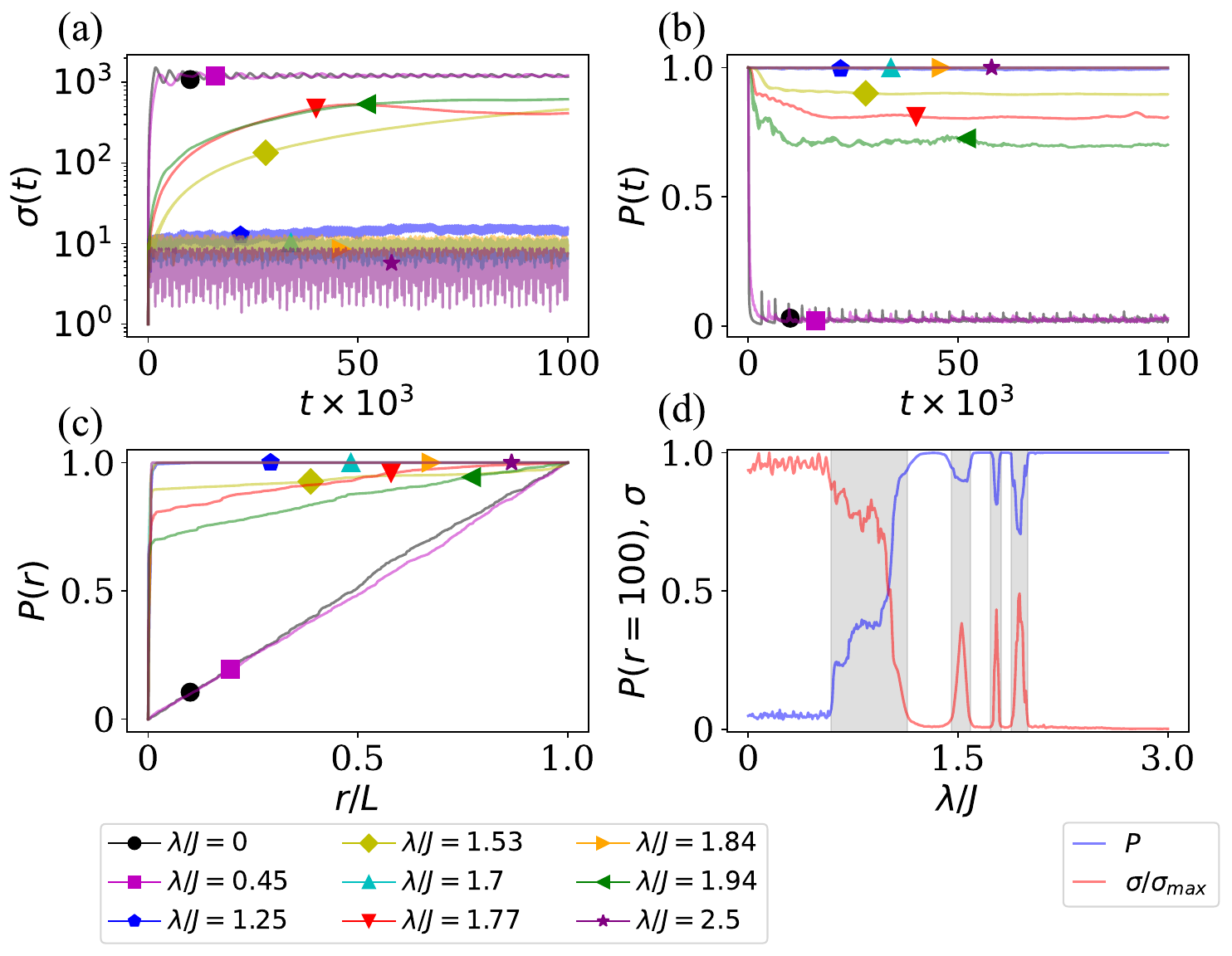}
     \caption{(a) $\sigma(t)$ (b) $P(r=100,\ t)$ for $\lambda/J=0,\ 0.45,\ 1.25,\ 1.53,\ 1.7,\ 1.77,\ 1.84,\ 1.95,\ 2.5$. (c) $P(r,\ t=10^5)$ as function of $r/L$. (d) The time evolved values of $\sigma(t=10^5)$ and $P(r=100,\ t=10^5)$ versus quasiperiodic disorder strength $\lambda/J$. The shaded regions represent the critical area in panel (d). For all cases, a system size of $L= 4181$ and periodic potential strength $V/J=0.875$ is considered. }
     \label{fig4}
\end{figure}

Furthermore, we compute two experimentally accessible quantities: the root mean square displacement $\sigma(t)$ and the survival probability $P(r,t)$ in Fig.~\ref{fig4}. The root mean square displacement is defined as
\begin{equation}
	\sigma(t) = \sqrt{\sum_{i=1}^L(i-i_0)^2|\psi_i(t)^2|}, 
\end{equation}
where $i_0$ is the initial position of the particle, chosen as $L/2$ in our calculations. And $|\psi(t)\rangle$ denotes the time-evolved wave function of the initial state, satisfying $|\psi(t)\rangle = e^{-i\hat{\mathcal{H}}t}|\psi_0\rangle$ and $|\psi_0\rangle=\hat{c}_{i_0}^{\dagger}|0\rangle$, which corresponds to a state localized on a single lattice site. $\sigma(t)$ characterizes the weighted average distance that a particle deviates from its initial position over time and as recently observed in quantum gas experiments \cite{PhysRevLett.120.160404}. For extended states, $\sigma(t)$ increases rapidly with time until saturation; for localized states, $\sigma(t)$ remains nearly unchanged; and for systems in the critical regime, $\sigma(t)$ also increases with time but at a slower rate compared to extended states. The survival probability is defined as 
\begin{equation}
	P(r,t) = \sum_{i = i_0-r/2}^{i_0+r/2} |\psi_i(t)|^2,
\end{equation}
which gives the probability of finding the particle within the interval $[i_0-r/2,\, i_0+r/2]$ at time $t$ \cite{PhysRevLett.123.025301, PhysRevB.97.060303, Xu_2020}. For extended states, since the particle rapidly departs from its initial position, $P(r=100, t)$ rapidly decays to zero as time increases. For localized states, $P(r=100, t)$ remains close to unity, while for critical states, due to the small but finite escape probability, $P(r=100, t)$ within a finite range $r$ still decreases over time, but at a significantly slower rate compared to extended states and eventually saturates at a finite value. The quantity $P(r, t=10^5)$ characterizes the distribution of particles after a long time of evolution. In the extended phase, the particle wave function becomes widely spread and the probability of finding the particle at all sites is almost equal to $1/L$, $P(r, t=10^5)$ returns to the order of unity only when $r/L \sim 1$. In contrast, for localized states, $P(r, t=10^5)\sim 1$ remains nearly independent of $r$ as expected. And for systems in the critical regime, the behavior lies between these two extremes. 

In our calculations, we choose $\lambda/J = 0$, $0.45$, $1.25$, $1.53$, $1.70$, $1.77$, $1.84$, $1.95$, and $2.5$. Fig.~\ref{fig4}(a) shows the time evolution of $\sigma(t)$ on a logarithmic scale. For localized cases with $\lambda/J = 1.25$, $1.70$, $1.84$, and $2.5$, $\sigma(t)$ remains close to zero throughout the evolution. For extended cases with $\lambda/J = 0$ and $0.45$, $\sigma(t)$ increases rapidly and then saturates. In contrast, for the critical cases at $\lambda/J = 1.53$, $1.77$ and $1.94$, $\sigma(t)$ grows with time and gradually approaches saturation at a slower rate than for extended cases. Similarly, as shown in Fig.~\ref{fig4}(b), for extended states, $P(r=100, t)$ rapidly decays to zero, while for localized states, $P(r=100, t)$ remains nearly constant around unity without noticeable temporal variation. For critical states, $P(r=100, t)$ lies between these two behaviors. In Fig.~\ref{fig4}(c), one can see that for extended states, $P(r, t=10^5)$ gradually approaches unity as $L$ increases, whereas for localized states, $P(r, t=10^5)$ quickly returns to and remains close to unity. For critical states, $P(r=100, t)$ again exhibits an intermediate behavior between the extended and localized limits. 

Compared with Fig.~\ref{fig3}, where the multiple reentrant localization transitions of the system are identified using eigenstate information, we can also observe these transitions through experimentally accessible quantities $\sigma$ and $P$, as shown in Fig.~\ref{fig4}(d). The results of $\sigma$ and $P$ still reveal the quasiperiodic disorder strength intervals for reentrant localization transitions, namely $1.45 \sim 1.59$, $1.73 \sim 1.81$ and $1.88 \sim 2.00$, nearly consistent with Fig.~\ref{fig3}(b). However, it should be noted that the first critical interval indicated in Fig.~\ref{fig4}(d), the first shaded region that extends $0.59 \sim 1.14$, is broader than the corresponding interval $0.99 \sim 1.14$ in Fig.~\ref{fig3}(b). This discrepancy arises because the mean $\mathcal{I}$ and $\mathcal{N}$ in Fig.~\ref{fig3}(b) are calculated using only a subset of eigenstates, whereas the time evolution considered here necessarily involves all eigenstates. 

\begin{figure}
     \centering
     \includegraphics[width=0.48\textwidth]{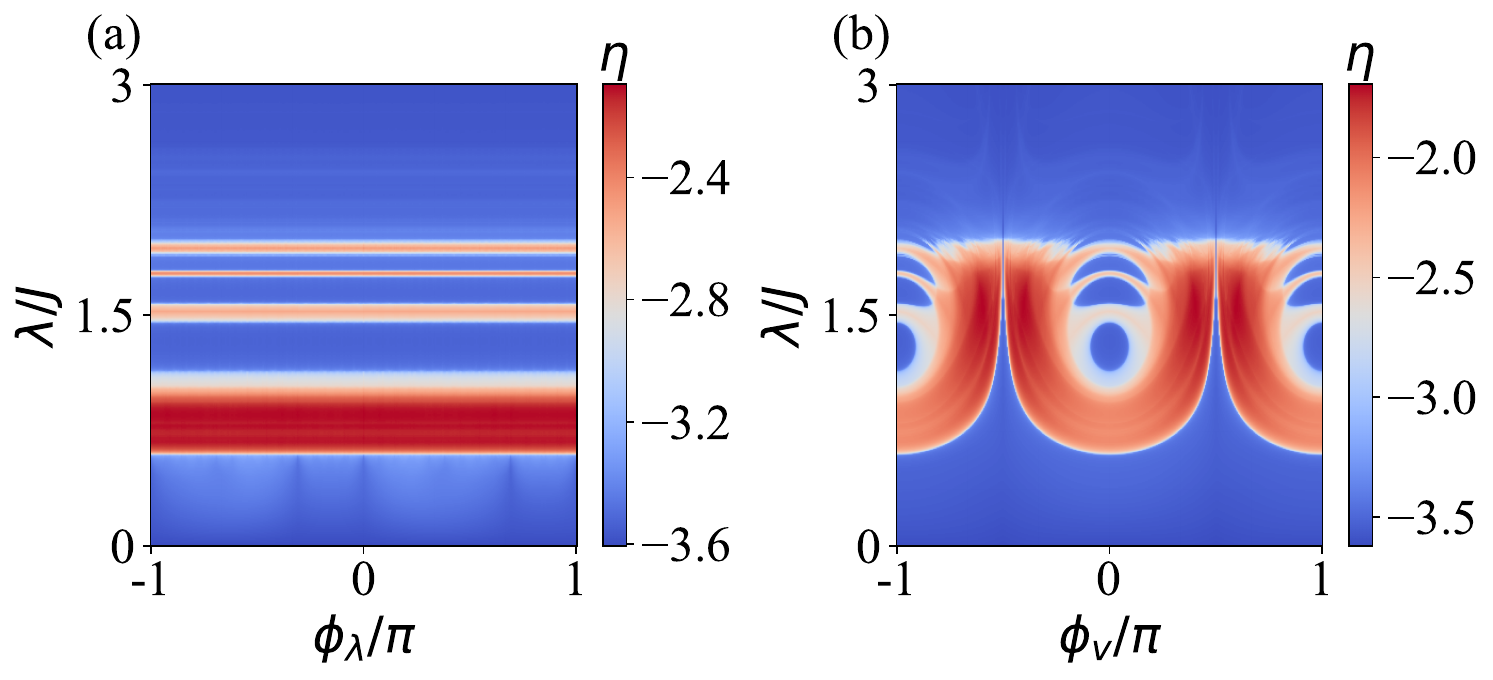}
     \caption{(a) $\eta$ phase for different $\phi_{\lambda}$ and quasiperiodic disorder strength $\lambda$ with $\phi_{v}=0$. (b) $\eta$ phase for different $\phi_{v}$ and quasiperiodic disorder strength $\lambda$ with $\phi_{\lambda}=0$. And for all cases, a system size of $L= 4181$ and periodic potential strength $V/J=0.875$ is considered. }
     \label{fig5}
\end{figure}

As shown in Fig.~\ref{fig2}, different initial phases have a significant impact on the localization transitions of the system. Next, we further explore the influence of phases on the multiple reentrant localization transitions. Fig.~\ref{fig5}(a) shows the behavior of $\eta$ as a function of the strength of the disorder $\lambda$ for different quasiperiodic disorder phases $\phi_{\lambda}$, with the periodic potential phase fixed at $\phi_v = 0$. It can be seen that as $\lambda$ increases, the system undergoes clear multiple reentrant localization transitions. However, these transitions are unaffected by variations in $\phi_{\lambda}$, consistent with the standard AAH model, in which the phase of the quasiperiodic disorder does not significantly influence the localization properties \cite{AAH}. In Fig.~\ref{fig5}(b), we investigate the effect of the periodic potential phase $\phi_v$ on the localization transitions. Without loss of generality, we set $\phi_{\lambda} = 0$. It is evident that the variation of $\phi_v$ has a pronounced impact on the multiple reentrant localization transitions. In the vicinity of $\phi_v = 0$, pronounced multiple reentrant localization transitions can be observed. And at $\phi_v = \pi/2 + n\pi$ with integer $n$, the model reduces to the standard AAH model, and no critical states are present. Moreover, the phase diagram of $\eta$ exhibits a periodic dependence on $\phi_v$ with a period of $\pi$, which is consistent with our choice of periodic potential frequency $\alpha = 1/2$. Compared with Ref. \cite{PhysRevB.111.134204}, where localization transitions appear only within a narrow range of $\phi_v$, here multiple reentrant localization transitions can be found over a much broader interval of $\phi_v$. 

\begin{figure}
     \raggedright
     \includegraphics[width=0.48\textwidth]{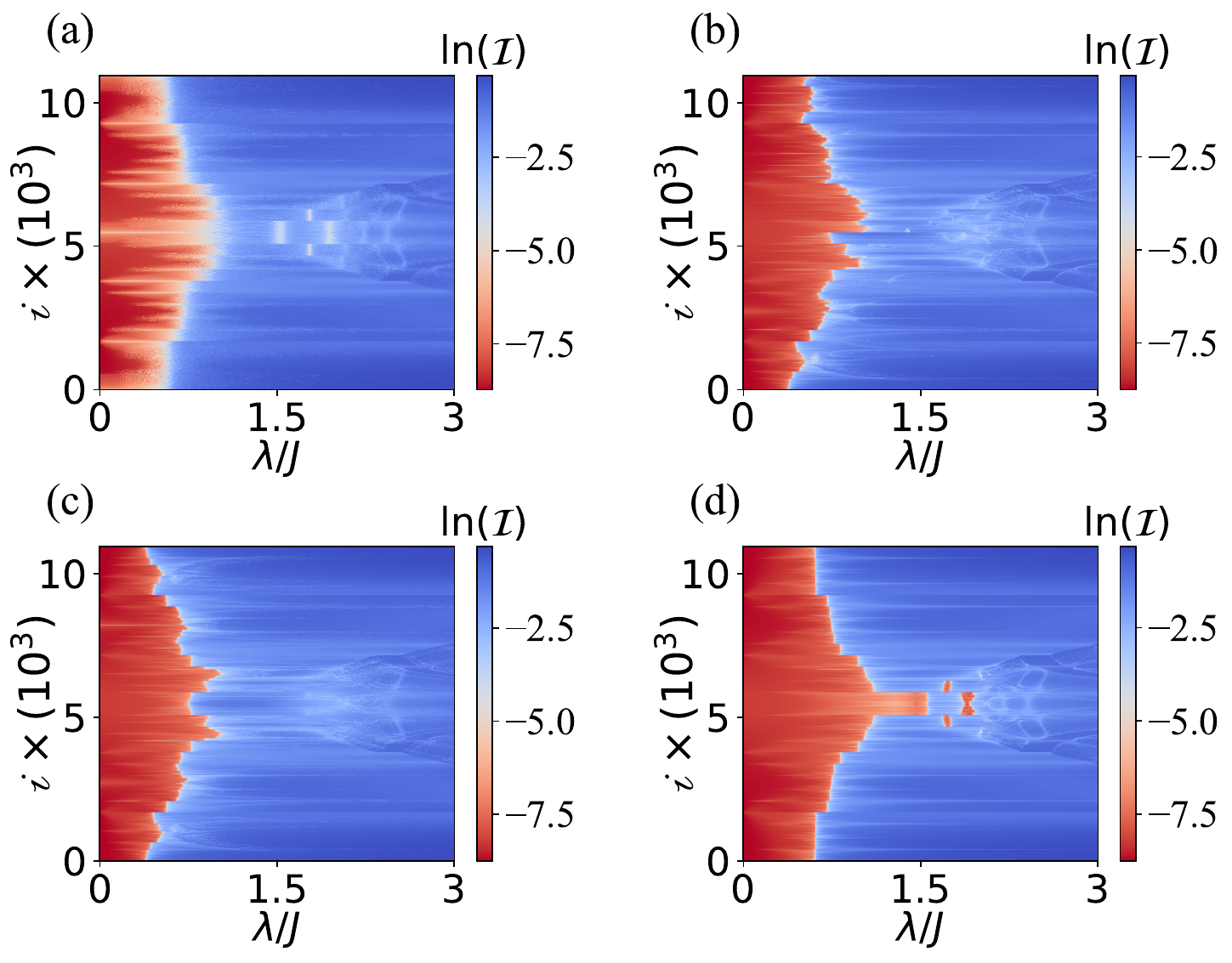}
     \caption{(a)–(d) depict the spectra for different quasiperiodic disorder strength $\lambda/J$ and indices of eigenstate potential strength under different perturbation, with system size $L=10946$ and periodic potential strength $V/J=0.875$. Specifically, (a) with perturbation term $\Delta_{\lambda} \frac{1}{L}\sum_i\left\lfloor\frac{i}{2}\right\rfloor\hat{n}_i$, which explicitly breaks the spatial translational symmetry. (b) with perturbation term $\Delta_{\lambda}\sum_{i\ mod\ 4=1}\hat{n}_i$. (c) with perturbation term $\Delta_{\lambda}(\sum_{i\ mod\ 4=1}\hat{n}_i-\sum_{i\ mod\ 4=0}\hat{n}_i)$, and (d) with perturbation term $\Delta_{\lambda}\sum_{i\ mod\ 2=1}\hat{n}_i$. $L=10946$, $V/J=0.875$, $\Delta_{\lambda}=0.05$, $\phi_{\lambda}=\phi_v=0$, and the colors represent the values of $\ln(\mathcal{I})$. }
     \label{fig6}
\end{figure}

To further investigate the robustness of the multiple reentrant localization phenomena, we examine four types of perturbation with a fixed perturbation strength $\Delta_{\lambda}=0.05$ at $V/J=0.875$ and $\phi_v=\phi_\lambda=0$, as shown in Fig.~\ref{fig6}. In Fig.~\ref{fig6}(a), the perturbation takes the form $\Delta_{\lambda}\,\frac{1}{L}\sum_i \left\lfloor \tfrac{i}{2} \right\rfloor \hat{n}_i$, where $\lfloor \cdot \rfloor$ denotes the floor function. This perturbation breaks the translational symmetry between neighboring dimer cells, but due to the normalization factor $1/L$, the perturbation difference between adjacent dimer unit cells remains very small, and the system can still be approximately regarded as consisting of dimer unit cells placed in a weak uniform electric field. In this case, the multiple reentrant localization transitions are slightly weakened but remain clearly visible, as illustrated in Fig.~\ref{fig6}(a). 

In Fig.~\ref{fig6}(b), $\Delta_{\lambda}\sum_{i\ mod\ 4=1}\hat{n}_i$, i.e., the perturbation acts on lattice sites with indices $1, 5, 9, \ldots$. This perturbation enlarges the unit cell to four sites and breaks the mirror symmetry of the original dimer cell. Under this perturbation, the multiple reentrant localization phenomena disappear completely as illustrated in Fig.~\ref{fig6}(b). 

To restore the mirror symmetry of the original dimer cell in the enlarged unit cell, we consider a modified perturbation of the form $\Delta_{\lambda}(\sum_{i\ mod\ 4=1}\hat{n}_i-\sum_{i\ mod\ 4=0}\hat{n}_i)$, which applies a $+\Delta_{\lambda}$ shift to sites $1, 5, 9, \ldots$, and a $-\Delta_{\lambda}$ shift to sites $4, 8, 12, \ldots$. while leaving all other sites unaffected. This construction ensures that the four-site unit cell retains the mirror symmetry of the original dimer. However, as shown in Fig.~\ref{fig6}(c), the multiple reentrant localization transitions again vanish, even under such a weak perturbation with $\Delta_{\lambda}=0.05$. 

Finally, in Fig.~\ref{fig6}(d), the perturbation $\Delta_{\lambda}\sum_{i\ mod\ 2=1}\hat{n}_i$, preserves the dimer unit cell structure, simply shifting the first site within each dimer by $+\Delta_{\lambda}$. In this case, the multiple reentrant localization transitions remain robust. These results highlight that the emergence of multiple reentrant localization is fundamentally a consequence of the competition between the dimerization effect and the quasiperiodic disorder modulation and that this competition crucially requires the preservation of the dimer unit cell. 

\begin{figure}
     \raggedright
     \includegraphics[width=0.48\textwidth]{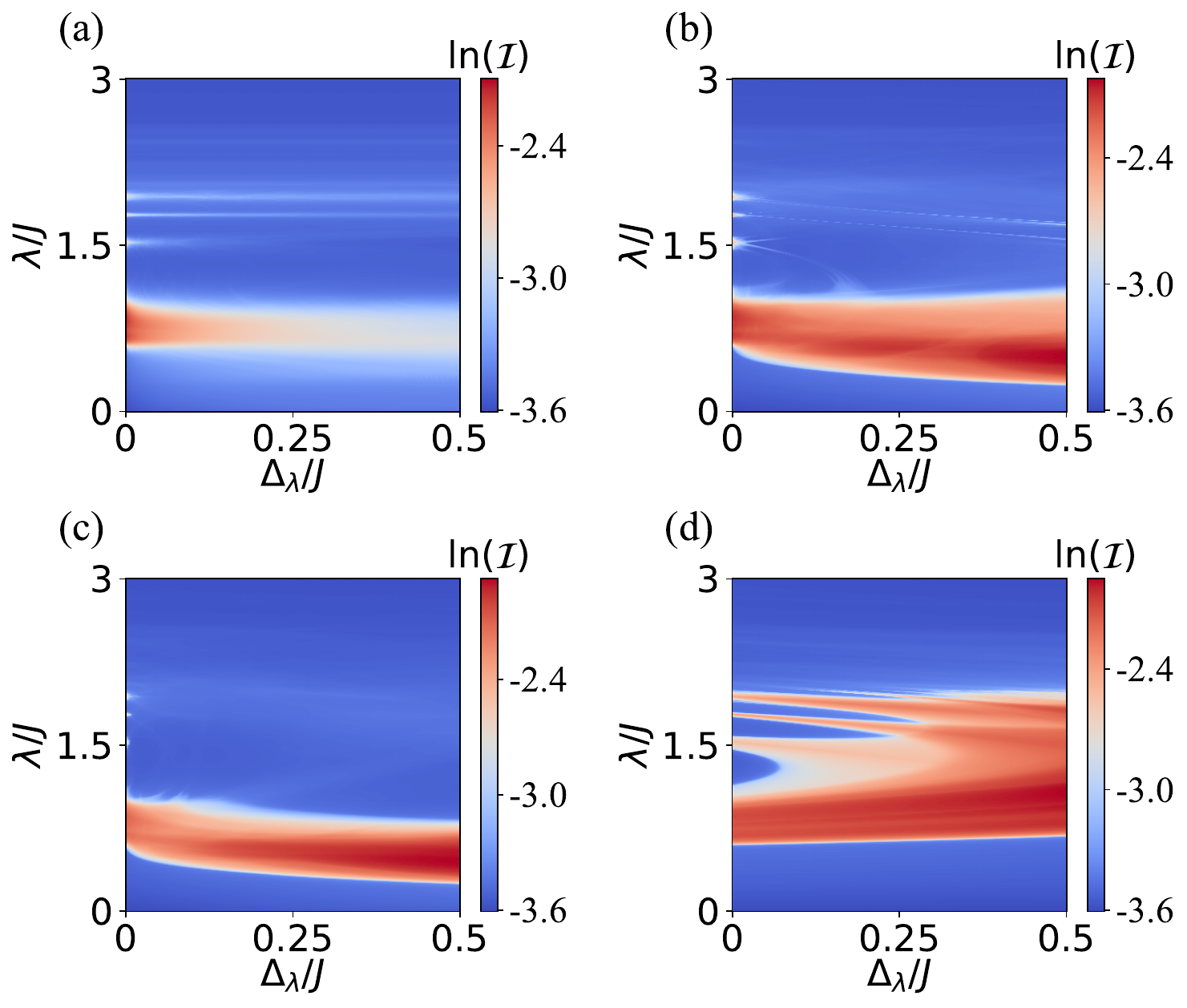}
     \caption{(a)–(d) show the $\eta$ as a function of the perturbation strength $\Delta/J$ and the quasiperiodic disorder strength $\lambda/J$, under different types of perturbations. Perturbation term with (a) $\Delta_{\lambda} \frac{1}{L}\sum_i\left\lfloor\frac{i}{2}\right\rfloor\hat{n}_i$, (b) $\Delta_{\lambda}\sum_{i\ mod\ 4=1}\hat{n}_i$, (c) $\Delta_{\lambda}(\sum_{i\ mod\ 4=1}\hat{n}_i-\sum_{i\ mod\ 4=0}\hat{n}_i)$ and (d) $\Delta_{\lambda}\sum_{i\ mod\ 2=1}\hat{n}_i$. $L=4181$, $V/J=0.875$, and $\phi_{\lambda}=\phi_v=0$. }
     \label{fig7}
\end{figure}

To further quantify the stability of these phenomena, we gradually increase the perturbation strength $\Delta_{\lambda}$ from zero and monitor the localization behavior, as shown in Fig.~\ref{fig7}. For the weak electric field–like perturbation in Fig.~\ref{fig7}(a), the multiple reentrant localization transitions are weakened with increasing $\Delta_{\lambda}$ but remain stable over a finite parameter range. In contrast, the perturbations leading to the four-site unit cell in Fig.~\ref{fig7}(b) and Fig.~\ref{fig7}(c), regardless of whether mirror symmetry is preserved, both rapidly suppress the multiple reentrant localization transitions as $\Delta_{\lambda}$ increases. In stark contrast, the dimer-preserving perturbation in Fig.~\ref{fig7}(d) continues to support robust multiple reentrant localization over a finite interval of perturbation strength. These results collectively demonstrate that the multiple reentrant localization transitions crucially depend on the preservation of the dimer unit cell. The phenomenon should thus be understood as a manifestation of the competition between the onsite periodic potential inherent to the dimer structure and the quasiperiodic disorder modulation.

\begin{figure*}
     \centering
     \includegraphics[width=\textwidth]{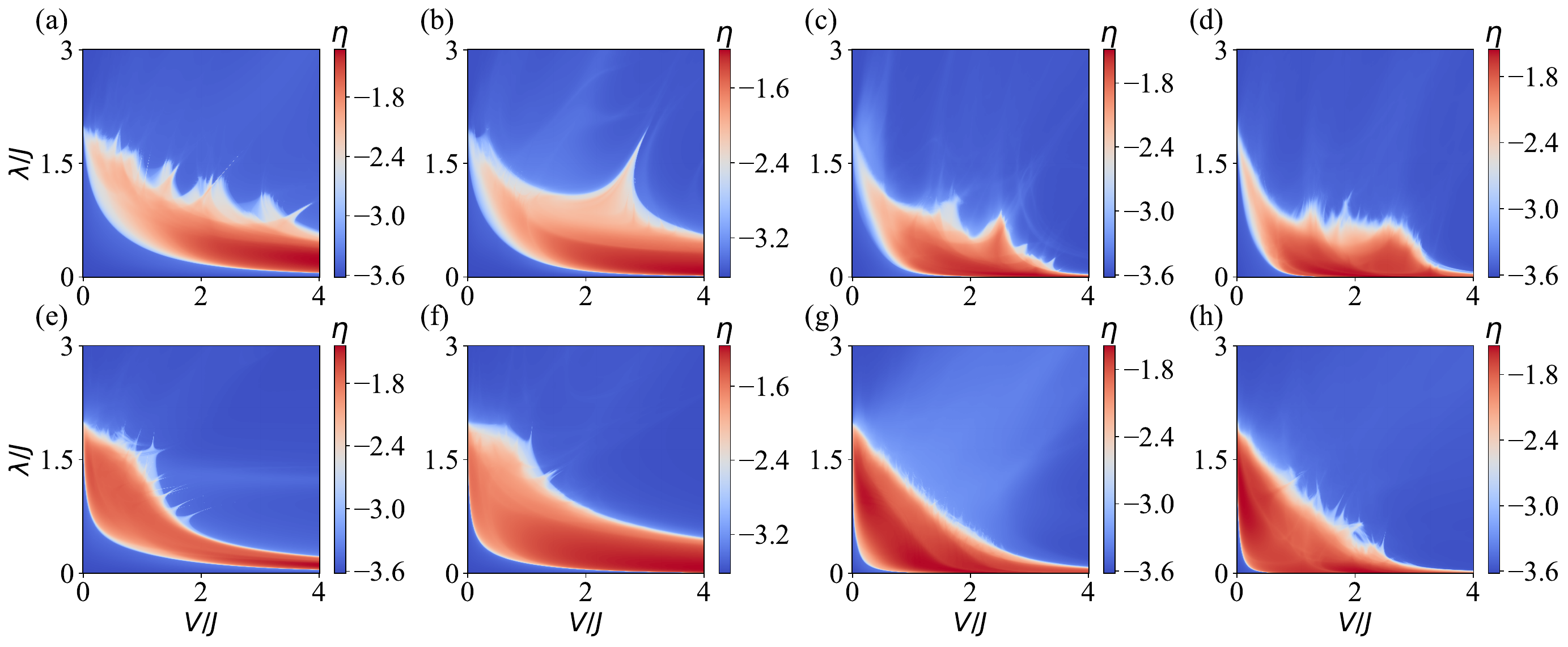}
     \caption{(a)–(h) show the distributions of $\eta$ as functions of the periodic potential strength $V/J$ and quasiperiodic disorder strength $\lambda/J$ with different frequency of the periodic potential $\alpha$ and frequency of the quasiperiodic disorder potential $\beta$. (a) and (e) $\alpha=\frac{1}{3}$, (b) and (f) $\alpha=\frac{1}{4}$, (c) and (g) $\alpha=\frac{1}{7}$, (d) and (h) $\alpha=\frac{1}{9}$. And the upper row, panels (a)–(d), corresponds to $\beta=\frac{\sqrt{5}-1}{2}$, while the lower row, panels (e)–(h), corresponds to $\beta=\frac{\sqrt{5}-1}{8}$. System size $L=4181$, and $\phi_{\lambda}=\phi_v=0$. }
     \label{fig8}
\end{figure*}

Next, we compute phase diagrams $\eta$ for different periodic potential frequencies $\alpha = 1/3$, $1/4$, $1/7$, and $1/9$, as functions of the periodic potential strength $V$ and the quasiperiodic disorder strength $\lambda$, as shown in Fig.~\ref{fig8}. In the first row of Fig.~\ref{fig8}, the quasiperiodic frequency is fixed at $\beta = (\sqrt{5}-1)/2$, while in the second row it is chosen as $\beta = (\sqrt{5}-1)/8$.

For $\beta = (\sqrt{5}-1)/2$, clear signatures of reentrant localization can be observed at $\alpha=1/2$, as already demonstrated in Fig.~\ref{fig1}(a). When $\alpha=1/3$, reentrant localization remains visible but appears weaker compared to $\alpha=1/2$, as shown in Fig.~\ref{fig8}(a). For $\alpha=1/4$, the range of $V/J$ supporting reentrant localization becomes narrower, as illustrated in Fig.~\ref{fig8}(b). Furthermore, when $\alpha$ is further reduced to $1/7$ and $1/9$, no reentrant localization is observed in the $\eta$ phase diagrams Fig. ~\ref{fig8}(c) and Fig.~\ref{fig8}(d).

When the quasiperiodic frequency is changed to $\beta = (\sqrt{5}-1)/8$, pronounced multiple reentrant localization transitions emerge in the $\alpha=1/2$ phase diagram Fig.~\ref{fig1}(c). For $\alpha=1/3$, reentrant localization persists but the multiple reentrant localization transitions are absent Fig.~\ref{fig8}(e). At $\alpha=1/4$, only weak reentrant localization is visible Fig.~\ref{fig8}(f), while for $\alpha=1/7$ no such features are observed Fig.~\ref{fig8}(g). Interestingly, for $\alpha=1/9$, a faint reentrant localization signal reappears, but without accompanying multiple reentrant localization transitions, as shown in Fig.~\ref{fig8}(h). These results demonstrate a clear trend: as $\alpha$ decreases, the phenomenon of reentrant localization gradually weakens and eventually disappears, while multiple reentrant localization transitions are robustly observed only near $\alpha=1/2$. 

\begin{figure}
     \centering
     \includegraphics[width=0.5\textwidth]{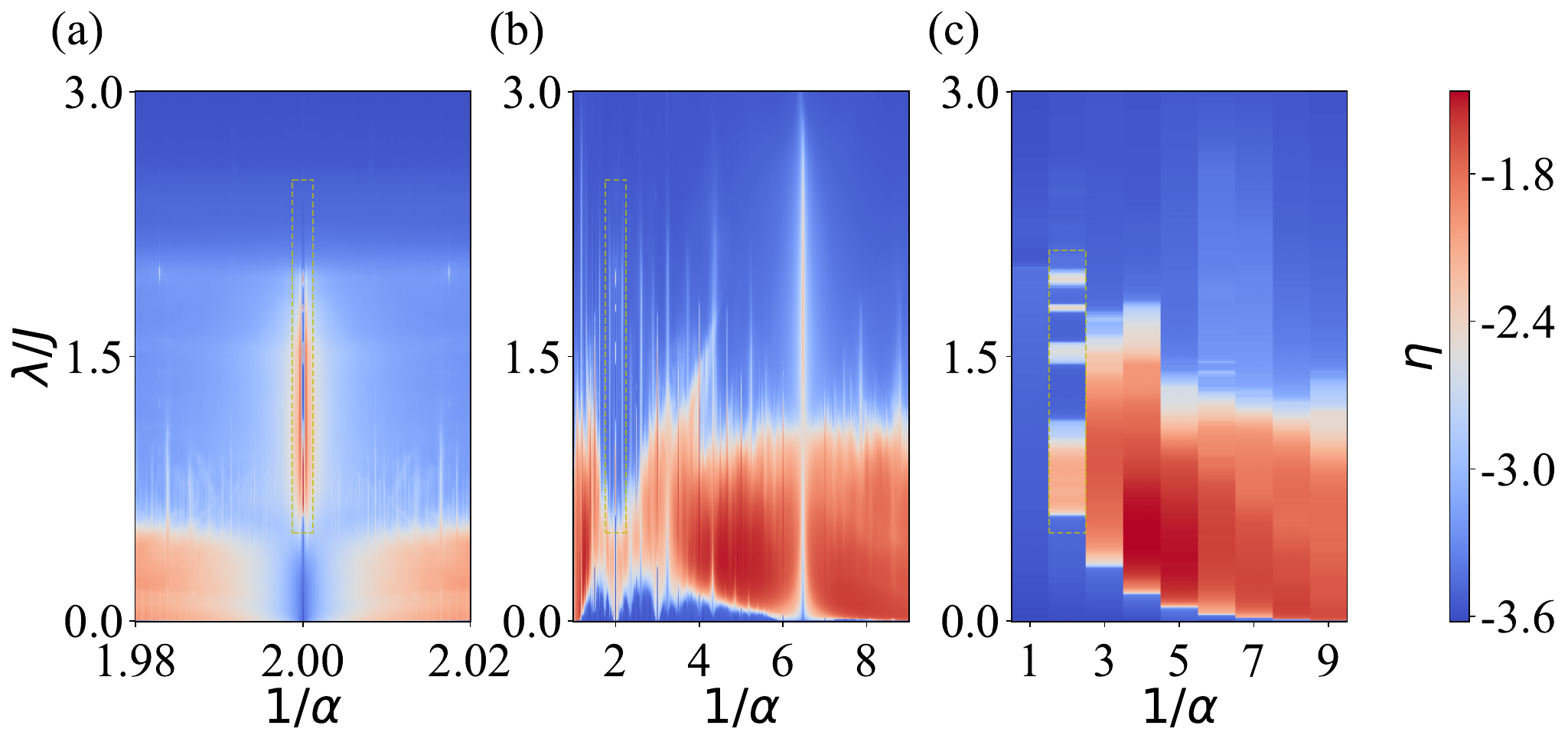}
     \caption{(a)–(c) show the $\eta$ as a function of the periodic potential period $\alpha$ and the quasiperiodic disorder strength $\lambda/J$, computed over different ranges of $\alpha$. In each panel, the yellow dashed box highlights the vicinity of $\alpha = 2$, where multiple reentrant localization features emerge. $L=4181$, $V/J=0.875$, and $\phi_{\lambda}=\phi_v=0$. The irregularity of the patterns arises in (b) from the fact that, in most cases, $\alpha$ cannot be taken as the reciprocal of an integer, thereby destroying the perfectly periodic potential of the $V$ term. }
     \label{fig9}
\end{figure}

In the parameter corresponding to the yellow vertical dashed line in Fig.~\ref{fig1}(c), we investigate the effect of varying $\alpha$ in the vicinity of $\alpha=1/2$. The results are presented in Fig.~\ref{fig9}(a), where the yellow dashed box highlights the multiple reentrant localization transitions that appear only when the system forms a dimer unit cell. Once $\alpha$ deviates from $1/2$, no signatures of multiple reentrant localization can be observed, indicating that even small deviations from the dimer configuration are sufficient to destroy the phenomenon. 

We further extend the calculation to a broader neighborhood of $\alpha$, with $\alpha \in [1/9, 1]$, as shown in Fig.~\ref{fig9}(b). Continually, multiple reentrant localization transitions are observed exclusively at $\alpha=1/2$, while no such features appear when $\alpha$ departs from the dimer condition. The phase diagram in Fig.~\ref{fig9}(b) appears rather irregular, because for most values of $\alpha$ that are not simple integer ratios, the periodic potential effectively acquires a very long period. At the system sizes accessible in our calculation, such long periods mimic quasiperiodic disorder and obscure clear patterns. To clarify this effect, we extract only those cases where $\alpha$ takes simple integer ratios, as shown in Fig.~\ref{fig9}(c). In complete agreement with Fig.~\ref{fig9}(b), multiple reentrant localization transitions are found only at the dimer point, $\alpha=1/2$. 

\section{Conclusion}\label{se4}

In conclusion, we have systematically investigated the localization properties of a one-dimensional system with a periodic potential and a quasiperiodic modulation. Using eigenstate-based measures, including the inverse participation ratio (IPR) $\mathcal{I}$, the normalized participation ratio (NPR) $\mathcal{N}$, and the averaged scaling parameter $\eta$, along with dynamical observables such as the root mean square displacement $\sigma(t)$ and the survival probability $P(r,t)$, we have constructed a comprehensive characterization of localization transitions in this model. In particular, we identify multiple reentrant localization transitions that are consistently captured by both eigenstate and dynamical analyses.

We demonstrate that the occurrence of multiple reentrant localization transitions crucially depends on the dimer structure of the unit cell, corresponding to a periodic frequency $\alpha = 1/2$. The dimer configuration ensures strong competition between the onsite periodic potential and the quasiperiodic disorder, which stabilizes these transitions. Perturbations that preserve the dimer unit cell allow multiple reentrant localization transitions to persist, whereas perturbations that modify the unit cell, such as by enlarging it, will destabilize the phenomenon of reentrant localization. Furthermore, varying $\alpha$ away from $1/2$ immediately suppresses multiple reentrant localization transitions, highlighting the unique role of the dimer condition. 

By exploring different quasiperiodic frequencies $\beta$ and periodic frequencies $\alpha$, we find that while weak or single reentrant localization signatures can appear at certain commensurate values of $\alpha$, robust multiple reentrant localization transitions occur exclusively at $\alpha = 1/2$. This establishes the dimer unit cell as the essential ingredient underlying the phenomenon, with the observed multiple reentrant localization transitions arising from the delicate competition between the onsite periodic potential of the dimer and the quasiperiodic modulation. Our results not only advance the understanding of localization phenomena in quasiperiodic systems but also provide experimentally accessible dynamical signatures that can serve as probes for detecting multiple reentrant localization transitions. 

\begin{acknowledgments}
The authors thank Yu-Hang Lu and Qin-Xin Liu for their helpful discussions. This work is supported by the National Natural Science Foundation of China under Grants No. 11974205 and 61727801, the Key Research and Development Program of Guangdong province (2018B030325002), and the National Natural Science Foundation of China under Grant 62131002. 
\end{acknowledgments}

\bibliography{apssamp}

\end{document}